\documentclass[onecolumn,showpacs,preprintnumbers,pra,amsfonts,amsmath]{revtex4}
\usepackage{graphicx}
\usepackage{bm}

\begin{document}

\title{Mean-field optical bistability of two-level atoms in structured reservoirs}
\author{G. A. Prataviera$^{1}$, A. C. Yoshida$^{2}$, and S. S. Mizrahi$^{3}$}
\email{prataviera@usp.br, yoshida@pontal.ufu.br, salomon@df.ufscar.br}
\affiliation{$^{1}$ Departamento de Administra\c{c}{\~a}o, FEA-RP, Universidade de S{\~a}%
o Paulo, 14040-905, Ribeir\~{a}o Preto, SP, Brazil \\
$^{2}$ Faculdade de Ci{\^e}ncias Integradas do Pontal, Universidade Federal
de Uberl\^{a}ndia,  38304-402, Ituiutaba, MG, Brazil \\
$^{3}$Departamento de F\'{\i}sica, CCET, Universidade Federal de S\~ao Carlos, 
13565-905, S\~ao Carlos, SP, Brazil}
\date{\today}

\begin{abstract}
We consider $N$ driven two-level atoms interacting with a structured reservoir. 
By dressing the collective operators within a semiclassical approach, we derive 
a master equation and a mean-field single particle effective Hamiltonian. 
This Hamiltonian describes the optical bistability phenomenon occurring in the relation 
between an input electromagnetic field and the effective output generated by the N atoms. 
The dissipative part of the master equation and the effective single-particle Hamiltonian contain new terms due the
reservoir structure of modes. In plotting the output field amplitude and phase, 
for a structured reservoir, as function of the input amplitude, one verifies the bistable behavior in both. 
We illustrate our results for two structured reservoirs: one having a Lorentzian shape 
for the distribution of modes, and the second is modeled as a photonic band-gap structure. 
\end{abstract}

\maketitle





\section{Introduction}

Optical and dynamic properties of atoms coupled to dissipative
environments with a tailored density of modes have been a hot topic 
in current research \cite{ref1,ref2,ref3,ref4,ref5,ref6,scully,tanas,keitel1,ref7,kurizki,ref8,ref9}. 
Several interesting and potentially useful effects such as: (i) suppression of 
spontaneous emission \cite{ref1,ref2,ref9}; (ii) modifications in the resonance fluorescence and 
absorption spectra of strongly driven two-level atoms 
\cite{ref3,ref4,ref5,ref6,scully,tanas,keitel1,ref7}, (iii) amplification without inversion 
\cite{keitel1}, and (iv) the possibility of effective control of atomic states \cite{kurizki}, have been reported 
by considering the interaction of atoms with structured reservoirs.
In general, theoretical studies of dissipative dynamics in structured reservoirs has been essentially devoted
to systems of one or a few atoms. However, for a large number of atoms, one interesting phenomenon is the 
one related to the \textit{optical bistability} (OB) \cite{lugiato,gibbs, xiao}. 
It originates from a nonlinear relation
between the intensity of an input field and an effective output field emerging from a collection 
of two-level atoms. Graphically, one sees the occurrence of a S-shaped curve, that corresponds to the 
existence of two stationary stable states for the atomic system, thus allowing the use of 
the system as an optical switch \cite{xiao}. 

Since its prediction and observation, in the 1970's, the OB \cite{szoke,mccall,mc2} has been the object
of intense research due its theoretical and experimental usefulness for studying nonlinear \cite{c1,c2,c3,c4,c5}, 
together with far from equilibrium effects arising in complex systems \cite{d1,d2,d3}. The main motivation was 
the potential applications in the construction of optical devices (\cite{xiao} and references therein).
The basic standard successful description of the OB consists
in considering a system of homogeneously broadened two-level atoms driven by
a coherent resonant field \cite{boni1}. As a matter of fact, the phenomenon arises from to the interplay between (1) 
dissipation due to the presence of a reservoir responsible for the atomic decay, (2) feedback 
due to the mean-field many-atoms effects, and (3) pumping of the atomic sample by an external field, 
occurring simultaneously. In a more specific context, in Ref. \cite{john} the authors study the 
change in the bistable S-shaped curve occurring in the population 
difference in impurity two-level atoms in a pseudophotonic band gap background 
to an applied laser field. So, the structure of modes of the reservoir may influence 
significantly the behavior of the bistable system, as we are going to further explore in this paper.

We shall treat the problem of atomic bistability in a
two-level N-atom system interacting with structured reservoirs. 
Formally we develop our calculations using the semiclassical dressed atom
approach for the collective operators, extending the treatment considered in 
\cite{scully,tanas} for a single atom. We derive a master equation
for $K$ ($1<K<N$) atoms and for a dilute system we obtain an effective
single-particle nonlinear Hamiltonian for a representative particle. The
effective Hamiltonian and the dissipative terms in the master equation
contain additional terms that are absent when a structureless reservoir is
considered. The relation between input and output fields is
obtained and, in contrast with the case of a structureless reservoir, we
found that besides the output amplitude, also the phase presents a bistable behavior. 
So, this feature allows probing the presence of a reservoir having a non-flat structure of modes.
We illustrate our results for: (1) a reservoir having a Lorentzian shape for
the modes distribution, and (2) a reservoir displaying a photonic band-gap structure. 

The article is organized as follows: In Sec. II we introduce the Hamiltonian describing
the system of atoms plus fields. In Sec. III we obtain a master equation for the N atoms
interacting with a structured reservoir. In Sec. IV, a mean-field
approximation is developed and an effective single-particle Hamiltonian is
obtained. In Sec. V the relation between input and output fields in the stationary state
is obtained. In Sec VI we illustrate the bistable behavior considering two
illustrative models for the reservoir. Finally, in Sec. VII we present a summary and
conclusions. 
%
\section{Atoms-field system}
%
We consider $N$ two-level atoms, with transition frequency $\omega _{0}$,
interacting within the rotating wave approximation (RWA) with a laser,
assumed as a classical electromagnetic field of frequency $\omega _{L}$ and
with the electric component $E_{in}e^{i\varphi }$, having an arbitrary phase 
$\varphi $. Besides, the atoms interact with a reservoir at $0K$, 
made of a continuum of modes, which is responsible for
the atomic decay. The Hamiltonian of the whole system is given by%
\begin{equation}
H=H_{A}+H_{R}+V_{AR}  \label{ham}
\end{equation}%
where%
\begin{eqnarray}
H_{A} &=&\frac{\hbar \omega _{0}}{2}S_{0}+\hbar F\left( e^{i\omega
_{L}t}S_{-}+e^{-i\omega _{L}t}S_{+}\right) ,
\label{hamA} \\
H_{R} &=&\int d\omega \ \left( \hbar \omega \right) b^{+}(\omega )b(\omega ),
\label{hamR} \\
V_{AR} &=&\hbar \int d\omega \ g(\omega )\left[ b(\omega )S_{+}e^{i\varphi }+b^{+}(\omega)S_{-}e^{-i\varphi }\right] .  
\label{hamAR}
\end{eqnarray}%
The Hamiltonian (\ref{hamA}) represents the $N$ two-level atoms pumped by
the laser field, with coupling constant $F=\mu E_{in}$ (we assume all atoms
having the same atomic dipole moment $\mu $). Admitting that the size of the
atoms is much smaller than the laser wavelength, the two active levels of
the atoms are described by the collective operators 
\begin{equation}
S_{0}=\sum_{i=1}^{N}\sigma _{0}(i);\quad S_{\pm }=e^{\mp i\varphi
}\sum_{i=1}^{N}\sigma _{\pm }(i),  \label{opercol}
\end{equation}%
and, $\sigma _{0}(i)$ and $\sigma _{\pm }(i)$ are the Pauli operators for a
single particle satisfying the commutation relations of the $SU\left(2\right)$ algebra, 
$\left[ \sigma _{0}(i),\sigma _{\pm }(j)\right] =\pm 2\delta _{i,j}\sigma
_{\pm }(i)$ and $\left[ \sigma _{+}(i),\sigma _{-}(j)\right] =\delta
_{i,j}\sigma _{0}(i)$. We remind that $\sigma _{+}=\left\vert e\right\rangle
\left\langle g\right\vert $, $\sigma _{-}=\left\vert g\right\rangle
\left\langle e\right\vert $, $\sigma _{0}=\left[ \left\vert e\right\rangle
\left\langle e\right\vert -\left\vert g\right\rangle \left\langle
g\right\vert \right] /2$, where $\left\vert e\right\rangle $ and $\left\vert
g\right\rangle $ refer to the higher and lower energy levels, respectively,
while $\hbar \omega _{0}$ is energy difference between the levels. The
Hamiltonian (\ref{hamR}) represents the reservoir modes, where the operator $%
b(\omega )$ ($b^{+}(\omega )$) annihilates (creates) a quantum of frequency $%
\omega $, and both satisfy the bosonic commutation relations, $\left[
b(\omega ),b^{+}(\omega ^{\prime })\right] =\delta \left( \omega -\omega
^{\prime }\right) $. Finally, Hamiltonian (\ref{hamAR}) corresponds to the
coupling between the reservoir modes and atoms, and $g(\omega )$ is the
coupling parameter, assumed to be frequency dependent, that characterizes a
structured reservoir. Furthermore, we assume that the atomic system is quite
diluted such that we disregard the direct interaction between the atoms, so
they will correlate and feel each other indirectly, as an effect of their
coupling with the reservoir modes. We also consider an undepleted laser
field so its dynamics is not taken into account.

In a referential frame rotating at frequency $\omega_{L} $ and in the interaction picture, 
with respect to the reservoir modes, the Hamiltonian (\ref{ham}) becomes 
\begin{equation}
H=H_{0S}+V_{AR}(t),  \label{ham2}
\end{equation}%
where 
\begin{equation}
H_{0S}=\delta S_{0}+F(S_{-}+S_{+}),  \label{ham3}
\end{equation}%
and $\delta =(\omega_{0} -\omega _{L})/2$ is the detuning frequency between
atoms and laser field; and 
\begin{equation}
V_{AR}(t)=\int d\omega \ g(\omega )\left[ b(\omega )S_{+}e^{i\left( \omega _{L}-\omega \right) t+i\varphi}+h.c.\right] .  \label{int1}
\end{equation}%
is the interaction between atoms and the reservoir.

In order to derive a dynamical equation for the atomic system, usually a
master equation, we follow an approach considered in references \cite%
{scully,tanas}. Instead of considering the atomic decay process
independently of the driving field, one assumes a coupled atom-driving field
decay, i. e., a generic atom is dressed by the driving field and coupled to
the reservoir modes. Within this approach we define the semiclassical dressed
collective operators 
\begin{subequations}
\label{eq}
\begin{align}
& \tilde{{S}}_{0}=\frac{1}{\Delta }\left( \delta {S}_{0}+F({S}_{-}+{S}%
_{+})\right)  \label{eqa} \\
& \tilde{{S}}_{-}=\frac{1}{2\Delta }\left( (\delta +\Delta ){S}_{-}-F{S}%
_{0}+(\delta -\Delta ){S}_{+})\right)  \label{eqb} \\
& \tilde{{S}}_{+}=\frac{1}{2\Delta }\left( (\delta -\Delta ){S}_{-}-F{S}%
_{0}+(\delta +\Delta ){S}_{+})\right) ,  \label{eqc}
\end{align}%
that satisfy the very same commutation relation of the $SU\left( 2\right) $
algebra $[\tilde{{S}}_{0},\tilde{{S}}_{\pm }]=\pm 2\tilde{{S}}_{+}$, $[%
\tilde{{S}}_{+},\tilde{{S}}_{-}]=\tilde{{S}}_{0}$. For later convenience we
write Eqs. (\ref{eqa})-(\ref{eqc}) in short as 
\end{subequations}
\begin{equation}
\tilde{{S}}_{j}=\sum_{l=-1}^{1}c_{jl}{S}_{l},\quad j=-1,0,1  \label{trans}
\end{equation}%
where $c_{jl}$ are the entries of the matrix 
\begin{equation}
\mathbb{C}\left( F,\delta ,\Delta \right) =\frac{1}{2\Delta }\left( 
\begin{array}{ccc}
\delta +\Delta & -F & \delta -\Delta \\ 
2F & 2\delta & 2F \\ 
\delta -\Delta & -F & \delta +\Delta%
\end{array}%
\right) ,  \label{trans1}
\end{equation}%
and $\Delta =\sqrt{\delta ^{2}+F^{2}}$. Definition (\ref{trans}) is
invertible, so the operators ${S}_{\pm }$ are related to the dressed ones
through the relation%
\begin{equation}
{S}_{k}=\sum_{j=-1}^{1}\tilde{c}_{kj}{\tilde{S}}_{j}.  \label{trans2}
\end{equation}%
As $\mathbb{C}^{-1}\left( F\right) =\mathbb{C}\left(
-F\right) $, the entries of the matrix $\mathbb{\tilde{C}}\left(
F,\delta ,\Delta \right) $ are $\tilde{c}_{kj}\left( F\right)
=c_{kj}\left( -F\right) $, and the dependence on the other parameters remains
the same.

Defining the single index coefficients $C_{j}=\tilde{c}_{1j}=\tilde{c}%
_{-1,-j}$ (where $C_{j=\pm 1}=\left( \delta + j\Delta \right) /\left( 2\Delta
\right) ,C_{0}=F/2\Delta $), the interaction Hamiltonian (\ref{int1}) can be
written as 
\begin{equation}
V_{AR}(t)=\sum_{j=-1}^{1}C_{j}\int d\omega \ g(\omega )\left[ B_{j}\left(
t,\omega \right) {\tilde{S}}_{j}+B_{j}^{\dagger }\left( t,\omega \right) {%
\tilde{S}}_{-j}\right] ,  \label{intera}
\end{equation}%
where $B_{j}\left( t\right) ={b}(\omega )e^{-i(\omega -\omega
_{L}-2\Delta jt)t+i\varphi }$.
%
\section{$N$-atom master equation}
%
In order to describe the dynamical evolution of an $N$-atom system state we
begin with the general non-Markovian master equation in the interaction
picture \cite{milburn} 
\begin{equation}
\frac{d\rho _{I,N}(t)}{dt}=-\int_{0}^{t}d\tau \ \mathrm{Tr}_{\mathcal{R}}%
\left[ V_{AR}(t),\left[ V_{AR}(t-\tau ),\rho _{I,N}(t-\tau )\rho _{R}\right] %
\right] ,  \label{master0}
\end{equation}
for the evolution of the density operator $\rho _{I,N}(t)$ of $N$ atoms
coupled to an unperturbed reservoir in thermal equilibrium. As
usual, Eq. (\ref{master0}) is obtained by tracing over the reservoir degrees
of freedom, and $\rho _{R}$ is the reservoir density
operator. The index $I$ stands for interaction picture. Besides, Eq. (\ref%
{master0}) was obtained in the weak coupling approximation, by assuming that
the atomic system plus reservoir density operator factorizes
as $\rho _{S+R}(t)=\rho _{S}(t)\rho _{R}$, at any time.

By inserting the interaction term given by Eq. (\ref{intera}) in the master
equation (\ref{master0}) we obtain 
\begin{eqnarray}
\frac{d\rho _{I,N}(t)}{dt} &=&-\sum_{j=-1}^{1}\sum_{j^{\prime
}=-1}^{1}C_{j}C_{j^{\prime }}\int_{0}^{\infty }d\omega g(\omega
)\int_{0}^{\infty }d\omega ^{\prime }\ g(\omega ^{\prime })\int_{0}^{t}d\tau
\notag \\
&&\times \mathrm{Tr}_{\mathcal{R}}\left[ B_{j}\left( t,\omega \right) {%
\tilde{S}}_{j}+B_{j}^{\dagger }\left( t,\omega \right) {\tilde{S}}_{-j},%
\left[ B_{j^{\prime }}\left( t-\tau ,\omega \right) {\tilde{S}}_{j^{\prime
}}+B_{j^{\prime }}^{\dagger }\left( t-\tau ,\omega \right) {\tilde{S}}%
_{-j^{\prime }},\rho _{I,N}(t-\tau )\rho _{R}\right] \right] .
\label{eqmaster1}
\end{eqnarray}%
We note that the density operator of the $N$-atom system at time $t$ depends
on the density operator at the previous time $t-\tau $. So, at this point,
we invoke the Markov assumption by replacing $\rho _{I,N}(t-\tau )$ in Eq. (%
\ref{eqmaster1}) by $\rho _{I,N}(t)$ and extend the upper limit in the
integral to infinity. As the reservoir is assumed in a vacuum state ($0K$), within that
approximation the trace operations result in 
\begin{eqnarray}
\mathrm{Tr}_{\mathcal{R}}\{b(\omega )b^{\dagger }(\omega ^{\prime })\rho
_{R}\} &=&\delta (\omega -\omega ^{^{\prime }}), \\
\mathrm{Tr}_{\mathcal{R}}\{b^{\dagger }(\omega )b(\omega ^{\prime })\rho
_{R}\} &=&0,  \label{eq45}
\end{eqnarray}%
and Eq. (\ref{eqmaster1}) becomes 
\begin{equation}
\frac{d\rho _{I,N}(t)}{dt}=-\sum_{j,j^{\prime }=-1}^{1}C_{j}C_{j^{\prime
}}e^{2i\Delta (j+j\prime )t}\xi _{j^{\prime }}\left( \omega _{L}+2j\Delta
\right) \left( \left[ \tilde{S}_{j},\tilde{S}_{-j^{\prime }}\rho _{I,N}(t)%
\right] e^{2i\Delta (j-j\prime )t}-\left[ \tilde{S}_{-j},\rho _{I,N}(t)%
\tilde{S}_{j^{\prime }}\right] e^{-2i\Delta (j-j\prime )t}\right)
\label{eqmaster2}
\end{equation}%
where 
\begin{equation}
\xi _{j}\left( \omega _{L}+2j\Delta \right) =\int_{0}^{\infty }d\omega \
\left( g(\omega )\right) ^{2}\left[ \pi \delta (\omega -\omega _{L}-2j\Delta
)\pm i\mathcal{P}\frac{1}{\omega -\omega _{L}-2j\Delta }\right] ,
\label{cor1}
\end{equation}%
and $\mathcal{P}$ stands for principal part. The first term at right hand
side of Eq. (\ref{cor1}) is responsible by the atomic decay rate while the
second term gives a shift in the atomic frequencies. Usually, these shifts
are incorporated in the atomic frequencies, but since our aim is to study
the decay rates, in this work it will be neglected and we write 
\begin{equation}
\xi _{j}\simeq \pi g^{2}(\omega _{L}+2j\Delta )\equiv \frac{\Gamma _{j}}{2V}
\label{decay}
\end{equation}%
($j=-1,0,+1$), where we introduced $\Gamma _{j}$ as the decay rates that depend on the
driving field and $V$ is the volume of the atomic cell that contains the $N$
atoms.

Back to the Schr\"{o}dinger representation, the master equation for an $N$-atom
system becomes. 
\begin{align}
& \frac{d\rho _{N}(t)}{dt}=-i\left[ H_{0S},\rho _{N}(t)\right] -\frac{\tilde{%
\Gamma}_{+}}{2V}\left\{ [{S}_{+},{S}_{-}{\rho }_{N}]+H.c.\right\}  \notag \\
& -\frac{\tilde{\Gamma}_{0}}{2V}\left\{ [{S}_{+},{S}_{0}{\rho }%
_{N}]+h.c.\right\} -\frac{\tilde{\Gamma}_{-}}{2V}\left\{ [{S}_{+},{S}_{+}{%
\rho }_{N}]+ H.c.\right\} ,  \label{mast}
\end{align}%
where $H.c.$ means Hermitian conjugate, and the effective decay rates are
given by 
\begin{equation}
\left( 
\begin{array}{c}
\tilde{\Gamma}_{-} \\ 
\tilde{\Gamma}_{0} \\ 
\tilde{\Gamma}_{+}%
\end{array}%
\right) =\frac{1}{4\Delta ^{2}}\left( 
\begin{array}{ccc}
-F^{2} & 2F^{2} & -F^{2} \\ 
-(\Delta +\delta )F & 2\delta F & (\Delta -\delta )F \\ 
(\Delta +\delta )^{2} & 2F^{2} & (\Delta -\delta )^{2}%
\end{array}%
\right) \left( 
\begin{array}{c}
\Gamma _{-} \\ 
\Gamma _{0} \\ 
\Gamma _{+}%
\end{array}%
\right) .  \label{effectiverate}
\end{equation}

So, in a structured reservoir the $N$-atom system contains decay rate
parameters $\tilde{\Gamma}_{j}$ that have a dependence on the frequency and
the intensity of the laser field. In contrast, in a non structured
reservoir, $g(\omega )=g_{0}$, the decay rates are either constant or zero, $%
\tilde{\Gamma}_{+}=2\pi g^{2}=\Gamma $, $\tilde{\Gamma}_{0}=\tilde{\Gamma}%
_{-}=0$, and the master equation (\ref{mast}) reduces to the well know form 
\cite{milburn}
\begin{equation}
\frac{d\rho _{N}(t)}{dt}=-i\left[ H_{0S},\rho _{N}(t)\right] -\frac{\Gamma }{%
2V}\left\{ [{S}_{+},{S}_{-}{\rho }_{N}]+H.c.\right\} .  \label{mastercol}
\end{equation}
The extra terms, third and fourth, appearing in Eq. (\ref{mast}) look like the dissipative 
terms that appear due to a squezeed vacuum reservoir \cite{gardiner} 
and where obtained in the context of a single atom in \cite{tanas}.
%
\section{ Mean-field approximation and effective Hamiltonian}
%
Now we will treat the $N$-atom system as a quantum BBGKY hierarchy of
equations similar to that of classical kinetic theory \cite{liboff,mcquarrie}. 
We consider a subsystem constituted of $K$ atoms, with $%
K<N$. The equation of motion for the $K$ atoms density operator $\rho _{K}$
is obtained by calculating the trace over the remaining $K+1,K+2,...,N$ atoms degrees of
freedom in Eq. (\ref{mast}), so getting 
\begin{align}
\frac{d{\rho _{K}}}{dt}& =Tr_{K+1,...,N}\left( \frac{d{\rho _{N}}}{dt}%
\right) =-i\mathrm{Tr}_{K+1,...,N}\left\{ \delta \lbrack {S}_{0},{\rho }%
_{N}]+F[{S}_{+},{\rho }_{N}]+F[{S}_{-},{\rho }_{N}]\right\}  \notag \\
& -\frac{\tilde{\Gamma}_{+}}{2V}\left\{ \mathrm{Tr}_{K+1,...,N}\left( [{S}%
_{+},{S}_{-}{\rho }_{N}]\right) +Tr_{K+1,...,N}\left( [{\rho }_{N}{S}_{+},{S}%
_{-}]\right) \right\}  \notag \\
& -\frac{\tilde{\Gamma}_{0}}{2V}\left\{ \mathrm{Tr}_{K+1,...,N}\left( [{S}%
_{+},{S}_{0}{\rho }_{N}]\right) +Tr_{K+1,...,N}\left( [{\rho }_{N}{S}_{0},{S}%
_{-}]\right) \right\}  \notag \\
& -\frac{\tilde{\Gamma}_{-}}{2V}\left\{ \mathrm{Tr}_{K+1,...,N}\left( [{S}%
_{+},{S}_{+}{\rho }_{N}]\right) +Tr_{K+1,...,N}\left( [{\rho }_{N}{S}_{-},{S}%
_{-}]\right) \right\} .  \label{masterK}
\end{align}%
Using the sums (\ref{opercol}) in terms of the microscopic operators 
Eq. (\ref{masterK}) becomes 
\begin{align}
\frac{d{\rho _{K}}}{dt}& =-i\sum_{i=1}^{K}\left\{ [\delta {\sigma }%
_{0}(i)+Fe^{-i\varphi }{\sigma }_{+}(i)+Fe^{i\varphi }{\sigma }_{-}(i),{\rho 
}_{K}]\right\}  \notag \\
& -\frac{(N-K)}{V}\sum_{i=1}^{K}\left\{ \frac{\tilde{\Gamma}_{+}}{2}[{\sigma 
}_{+}(i),\mathrm{Tr}_{K+1}\left( {\sigma }_{-}(K+1){\rho }_{K+1}\right)
]\right.  \notag \\
& +\frac{\tilde{\Gamma}_{0}}{2}e^{-i\varphi }[{\sigma }_{+}(i),\mathrm{Tr}%
_{K+1}\left( {\sigma }_{0}(K+1){\rho }_{K+1}\right) ]  \notag \\
& \left. +\frac{\tilde{\Gamma}_{-}}{2}e^{-2i\varphi }[{\sigma }_{+}(i),%
\mathrm{Tr}_{K+1}\left( {\sigma }_{+}(K+1){\rho }_{K+1}\right) ]+H.c.\right\}
\notag \\
& -\sum_{i,i^{\prime }=1}^{K}\left\{ \left( \frac{\tilde{\Gamma}_{+}}{2}[{%
\sigma }_{+}(i),{\sigma }_{-}(i^{\prime }){\rho }_{K}]+\frac{\tilde{\Gamma}%
_{0}}{2}e^{-i\varphi }[{\sigma }_{+}(i),{\sigma }_{0}(i^{\prime }){\rho }%
_{K}]\right. \right.  \notag \\
& \left. \left. +\frac{\tilde{\Gamma}_{-}}{2}e^{-2i\varphi }[{\sigma }%
_{+}(i),{\sigma }_{+}(i^{\prime }){\rho }_{K}]\right) +H.c.\right\} .
\label{masterK2}
\end{align}%
For a single representative atom of the system ($K=1$), the equation of
motion for $\rho _{1}$ will depend on the two atoms density operator $\rho
_{2}$, and so on for the whole hierarchy, meaning that the equation of
motion for $\rho _{K}$ will depend on the state $\rho _{K+1}$. However, for a
dilute system the higher-order atomic correlations may be disregarded, and
this is achieved when one factorizes ${\rho }_{2}$ as $\rho _{1}\otimes \rho
_{1}$, so a generic atom is assumed to move in a mean-field produced by all
the other atoms, which is a kind of Hartree approximation. Implementing this
approximation and dropping the subscript in $\rho _{1}$, the equation (\ref%
{masterK2}) reduces to 
\begin{eqnarray}
\frac{d{\rho }}{dt}=&-& i[{H}_{ef}\left[ {\rho }\right] ,{\rho }]-\frac{%
\tilde{\Gamma}_{+}}{2V}\left( [{\sigma }_{+},{\sigma }_{-}{\rho }%
]+H.c.\right)  \notag \\
&-&\frac{\tilde{\Gamma}_{0}}{2V}\left( e^{-i\varphi }[{\sigma }_{+},{\sigma 
}_{0}{\rho }]+H.c.\right)  \notag \\
&-&\frac{\tilde{\Gamma}_{-}}{2V}\left( e^{-i2\varphi
}[{\sigma }_{+},{\sigma }_{+}{\rho }]+H.c.\right) ,  
\label{masq}
\end{eqnarray}%
where the single particle effective Hamiltonian is 
\begin{equation}
{H}_{ef}\left[ \rho \right] =\delta {\sigma }_{0}+\left\{ e^{-i\varphi }%
\left[ \mu E_{in}-i\frac{(N-1)}{V}\left( \frac{\tilde{\Gamma}_{-}}{2}\langle {\sigma }%
_{+}\rangle e^{-i\varphi }+\frac{\tilde{\Gamma}_{0}}{2}\langle {\sigma }%
_{0}\rangle +\frac{\tilde{\Gamma}_{+}}{2}\langle {\sigma }_{-}\rangle
e^{i\varphi }\right) \right] {\sigma }_{+}+H.c.\right\} ,  \label{hamil}
\end{equation}%
which contains nonlinear terms corresponding to a mean-field that is due to
the remaining $N-1$ atoms.

From the second term in the brackets in the Hamiltonian (\ref{hamil}) we see that
effectively a single generic atom is excited by the input field amplitude $%
E_{in}$ plus an extra polarization field density $ \epsilon^{*} _{pol}(t)$, where
\begin{equation}
\epsilon _{pol}(t)=i\frac{(N-1)}{2\mu V}e^{i\varphi }\left( \tilde{\Gamma}%
_{-}\langle {\sigma }_{-}\rangle e^{i\varphi }+\tilde{\Gamma}_{0}\langle {%
\sigma }_{0}\rangle +\tilde{\Gamma}_{+}\langle {\sigma }_{+}\rangle
e^{-i\varphi }\right)
\end{equation}%
originated from the other $(N-1)$ atoms that produce a mean-field effect,
and is proportional to the uniform atomic density in the cell $N/V$,
for $N\gg 1$. When the function $g^{2}(\omega )$ is assumed being frequency
independent (white noise limit) the polarization field reduces to the simple
expression%
\begin{equation}
\epsilon _{pol}(t)=i\frac{(N-1)}{2\mu V}\Gamma \langle {\sigma }_{+}\rangle .
\end{equation}

The equations of motion for the atomic operators mean values, derived from the 
master equation (\ref{masq}), are 
\begin{equation}
\frac{d}{dt}\langle \sigma _{0}\rangle =2i\mu \left( {\epsilon_{out} }(t)\langle
\sigma _{-}\rangle -{\epsilon_{out} }^{\ast }(t)\langle \sigma _{-}\rangle ^{\ast
}\right) +\tilde{\Gamma}_{0}\left( \langle {\sigma }_{+}\rangle e^{-i\varphi
}+\langle {\sigma }_{-}\rangle e^{i\varphi }\right) -\tilde{\Gamma}%
_{+}(1+\langle s_{0}\rangle ),  \label{s0}
\end{equation}%
\begin{equation}
\frac{d}{dt}\langle \sigma _{-}\rangle =-2i\delta \langle \sigma _{-}\rangle
+i\mu{\epsilon_{out} }^{\ast }(t)\langle \sigma _{0}\rangle +\frac{e^{-i\varphi }}{2}%
\left( \tilde{\Gamma}_{-}\langle \sigma _{+}\rangle e^{-i\varphi }-\tilde{%
\Gamma}_{+}\langle \sigma _{-}\rangle e^{i\varphi }\right) +\frac{\tilde{%
\Gamma}_{0}}{2}e^{-i\varphi },  \label{s-}
\end{equation}%
and $\langle \sigma _{+}\rangle =\langle \sigma _{-}\rangle ^{\ast }$. The
total effective output field transmitted from the sample is defined as 
\begin{eqnarray}
\epsilon _{out}(t) &=&E_{in}e^{i\varphi }+\epsilon _{pol}(t)  \notag \\
&=&E_{in}e^{i\varphi }+i\frac{(N-1)}{2\mu V}\tilde{\Gamma}_{+}\langle {%
\sigma }_{+}\rangle +i\frac{(N-1)}{2\mu V}e^{i\varphi }\left( \tilde{\Gamma}%
_{-}\langle {\sigma }_{-}\rangle e^{i\varphi }+\tilde{\Gamma}_{0}\langle {%
\sigma }_{0}\rangle \right) .  \label{campototal}
\end{eqnarray}%
When $g(\omega )=g_{0}$, in Eq. (\ref{campototal}), the only remaining term
is $\tilde{\Gamma}_{+}\rightarrow \Gamma $ (proportional to $\sigma _{+}$),
a constant, while the other terms vanish. The mean-field extra terms,
proportional to $\langle {\sigma }_{-}\rangle $ and $\langle {\sigma }%
_{0}\rangle $, are due to the structured reservoir. The Eq. (\ref{campototal}), for
the output field $\epsilon _{out}(t)$, is more inclusive than the others deduced
in Refs. \cite{boni1, d1, bergou1}, because the mean-field approximation
contributes with additional terms that are sensible to the mode-structured reservoir.
%
\section{Stationary solution and the input-output fields relation}
%
When the solutions of Eqs. (\ref{s0}) and (\ref{s-}) are inserted in Eq. (%
\ref{campototal}) we get the output field (total field) as a function of the
input one. Here we are interested to study the influence of the structured
reservoir on the bistable steady state output field amplitude as a function of
input field $E_{in}$. The stationary solutions of Eqs. (\ref{s0}) and (\ref%
{s-}) are obtained by setting $d\langle \sigma _{0}\rangle /dt=$ $d\langle
\sigma _{-}\rangle /dt=0$, resulting in 
\begin{equation}
\langle \sigma _{0}\rangle _{ss}=-\frac{\tilde{\Gamma}_{+}(16\delta ^{2}+%
\tilde{\Gamma}_{+}^{2}-\tilde{\Gamma}_{-}^{2})-2\tilde{\Gamma}_{0}^{2}(%
\tilde{\Gamma}_{+}+\tilde{\Gamma}_{-})-2\tilde{\Gamma}_{0}[i(-4i\delta +%
\tilde{\Gamma}_{+}+\tilde{\Gamma}_{-})\mu{\epsilon }_{ss}e^{-i\varphi }+c.c.]}{%
\tilde{\Gamma}_{+}(16\delta ^{2}+\tilde{\Gamma}_{+}^{2}-\tilde{\Gamma}%
_{-}^{2})+2\mu \tilde{\Gamma}_{0}[i(4i\delta +\tilde{\Gamma}_{+}+\tilde{\Gamma}%
_{-}) {\epsilon }_{ss}e^{-i\varphi }+c.c.]-4\mu^2 \tilde{\Gamma}_{-}\left(  {%
 \epsilon }_{ss}^{2}e^{-2i\varphi }+{c.c.}\right) +8\mu^2\tilde{\Gamma}%
_{+} \left\vert {\epsilon }_{ss}\right\vert ^{2}}  \label{stat1}
\end{equation}%
and%
\begin{equation}
e^{-i\varphi }\langle \sigma _{+}\rangle _{ss}=-\frac{%
\begin{array}{c}
\left\{ -\tilde{\Gamma}_{0}\tilde{\Gamma}_{+}\left( 4i\delta +\tilde{\Gamma}%
_{+}+\tilde{\Gamma}_{-}\right) +2i[\tilde{\Gamma}_{0}^{2}-\tilde{\Gamma}%
_{+}(4i\delta +\tilde{\Gamma}_{+})]\mu {\epsilon }_{ss}e^{-i\varphi }\right. \\ 
\left. +2i(\tilde{\Gamma}_{0}^{2}+\tilde{\Gamma}_{+}\tilde{\Gamma}_{-})\mu {%
\epsilon }_{ss}^{\ast }e^{i\varphi }-4\mu^2 \tilde{\Gamma}_{0}({\epsilon }%
_{ss}^{2}e^{-2i\varphi }+\left\vert {\epsilon }_{ss}\right\vert ^{2})\right\}%
\end{array}%
}{\tilde{\Gamma}_{+}(16\delta ^{2}+\tilde{\Gamma}_{+}^{2}-\tilde{\Gamma}%
_{-}^{2})+2\mu \tilde{\Gamma}_{0}[i(4i\delta +\tilde{\Gamma}_{+}+\tilde{\Gamma}%
_{-}) {\epsilon }_{ss}e^{-i\varphi }+c.c.]-4\mu^2 \tilde{\Gamma}_{-}\left( {%
\epsilon }_{ss}^{2}e^{-2i\varphi }+{c.c.}\right) +8\mu^2\tilde{\Gamma}%
_{+} \left\vert {\epsilon }_{ss}\right\vert ^{2}},  \label{stat2}
\end{equation}%
where $\epsilon _{ss}$ is the output field at the stationary state.

In particular, for a non-structured (ns) reservoir, $g(\omega )=g_{0}$, Eqs. (\ref%
{stat1}) and (\ref{stat2}) simplify to the well known results \cite{boni1} 
\begin{equation}
\langle \sigma _{0}\rangle _{ss}^{(ns)}=-\frac{(16\delta ^{2}+\Gamma ^{2})}{%
16\delta ^{2}+\Gamma ^{2}+8\mu^2 \left\vert {\epsilon }_{ss}\right\vert ^{2}}%
,\qquad e^{-i\varphi }\langle \sigma _{+}\rangle _{ss}^{(ns)}=\frac{2(-4\delta
+i\Gamma )\mu {\epsilon }_{ss}e^{-i\varphi }}{16\delta ^{2}+\Gamma
^{2}+8\mu^2 \left\vert {\epsilon }_{ss}\right\vert ^{2}}.
\end{equation}%
for a single atom pumped by an external field ${\epsilon }%
_{ss}=E_{in}e^{i\varphi }+i\left( (N-1)/2\mu V\right) \Gamma \langle {\sigma 
}_{+}\rangle _{ss}$, which is due to a collective effect
produced by the mean-field. For a structured reservoir the stationary solution
changes in an essential way: besides of the terms proportional to $%
\left\vert {\epsilon }_{ss}\right\vert ^{2}$, others terms, proportional to $%
{\epsilon }_{ss}$ and ${\epsilon }_{ss}^{2}$, appear additionally.
These terms show some similarity to those produced by the decay in a
squeezed vacuum \cite{haas1,gardiner,bergou1,leonardo}, without mean-field effects, as
reported in \cite{tanas} in the case of a single atom. 

Writing ${\epsilon}_{ss}=E_{ss}e^{i\theta }$
(with the explicit introduction of a phase $\theta $) in the stationary state of Eq. (\ref{campototal}), we obtain the following
relation between the input and output fields 
\begin{eqnarray}
E_{in}^{2} &=&\left( E_{ss}\cos \Phi +\frac{\left( N-1\right) }{2V\mu }%
\left( \tilde{\Gamma}_{+}-\tilde{\Gamma}_{-}\right) \rm{Im} \left( e^{-i\varphi
}\langle \sigma _{+}\rangle _{ss}\right) \right) ^{2}  \notag \\
&&+\left( E_{ss}\sin \Phi -\frac{\left( N-1\right) }{2V\mu }\left( \tilde{%
\Gamma}_{+}+\tilde{\Gamma}_{-}\right) \left[ \rm{Re} \left( e^{-i\varphi }\langle
\sigma _{+}\rangle _{ss}\right) +\tilde{\Gamma}_{0}\langle \sigma
_{0}\rangle _{ss}\right] \right) ^{2}  \label{stat11}
\end{eqnarray}%
where $\langle \sigma _{+}\rangle _{ss}$ and $\langle
\sigma_{0}\rangle _{ss}$, from Eqs. (\ref{stat1}) and (\ref{stat2}), depend on the phase difference $\Phi =\theta
-\varphi $. We note that the nonlinear dependence of $\theta$ on $E_{in}$ is a manifestation of the
intrinsic frequency distribution of the reservoir. Thus, in order to determine graphically
the relations between the real and imaginary parts of $\epsilon_{ss}$ versus $E_{in}$, 
or $E_{ss}$ and $\Phi$ versus $E_{in}$, for each input value $E_{in}$, the 
two output dependent variables must satisfy Eq. (\ref{stat11})
for the unique independent variable $E_{in}$. 
In the white noise approximation we have $\tilde{\Gamma}_{-}=\tilde{\Gamma}_{0}=0$, 
$\tilde{\Gamma}_{+}=\Gamma $, so Eq. (\ref{stat11}) simplifies to the already 
known phase independent form \cite{xiao}, 
\begin{equation}
E_{in} =E_{ss}\left\{ \left( 1+\frac{\left( N-1\right) }{V \left(
1+16\left( \delta ^{2}/\Gamma ^{2}\right) +8\mu^2 E_{ss}^{2}/\Gamma ^{2}\right) }
\right) ^{2}+\left( \frac{4\left( N-1\right) \delta /\Gamma }{V \left(
1+16\left( \delta ^{2}/\Gamma ^{2}\right) +8\mu^2 E_{ss}^{2}/\Gamma ^{2}\right) }
\right) ^{2}\right\} ^{1/2}, 
\label{Ecampo}
\end{equation}
which displays the bistability phenomenon when output versus input fields are plotted. 
The first term in the braces of Eq. (\ref{Ecampo}) 
corresponds to the absorptive regime, when the atoms
are driven near resonance, while the second term stands for the dispersive
regime, when the atoms are driven far from resonance ($\delta /\Gamma \gg 1$) 
and nonlinear refractive effects dominate \cite{xiao}. Very characteristically, in this approximation 
(structureless reservoir), there is no phase dependence induced by the atoms. 
Therefore the presence of a phase shift indicates the existence of a structured reservoir (colored noise). 
%
\section{Structured reservoirs: output field amplitude and phase bistability}
%
In order to get a direct insight into the problem of bistability with a structured reservoir 
we consider the case when frequencies of the laser field and atomic transition are resonant ($\delta =0$).
Furthermore, calling ${\rm Re}\left( \epsilon _{out}\right) \equiv \epsilon _{x}$ and $
{\rm Im}\left( \epsilon _{out}\right) \equiv \epsilon _{y}$, Eq. (\ref{stat1}) becomes 

\begin{equation}
\left\langle \sigma _{0}\right\rangle _{ss} =\frac{-\left( \tilde{\Gamma}_{+}+\tilde{\Gamma}_{-}\right) \left[
\tilde{\Gamma}_{+}\left( \tilde{\Gamma}_{+}-\tilde{\Gamma}_{-}\right) -2%
\tilde{\Gamma}_{0}^{2}+4\mu\tilde{\Gamma}_{0}\epsilon _{y}\right] }{\left( 
\tilde{\Gamma}_{+}+\tilde{\Gamma}_{-}\right) \left[ \tilde{\Gamma}_{+}\left( 
\tilde{\Gamma}_{+}-\tilde{\Gamma}_{-}\right) -4\mu\tilde{\Gamma}_{0}\epsilon_{y}
+8\mu^2\epsilon _{y}^{2}\right] +8\mu^2\left( \tilde{\Gamma}_{+}-\tilde{\Gamma}%
_{-}\right) \epsilon _{x}^{2}}
\label{so}
\end{equation}
and the real and imaginary components of $\left\langle \sigma _{+}\right\rangle _{ss}$ are,  
\begin{equation}
{\rm Re}\left\langle \sigma _{+}\right\rangle _{ss}=\frac{\left( \tilde{\Gamma}_{0}-2 
\mu \epsilon _{y}\right) 
\tilde{\Gamma}_{+}\left( \tilde{\Gamma}_{+}+\tilde{\Gamma}_{-}\right) +8 \mu^2%
\tilde{\Gamma}_{0}\epsilon _{x}^{2}}{\left( \tilde{\Gamma}_{+}+\tilde{\Gamma}%
_{-}\right) \left[ \tilde{\Gamma}_{+}\left( \tilde{\Gamma}_{+}-\tilde{\Gamma}%
_{-}\right) -4\mu\tilde{\Gamma}_{0}\epsilon _{y}+8\mu^2\epsilon _{y}^{2}\right]
+8\mu^2\left( \tilde{\Gamma}_{+}-\tilde{\Gamma}_{-}\right) \epsilon _{x}^{2}},
\label{res+}
\end{equation}

\begin{equation}
{\rm Im}\left\langle \sigma _{+}\right\rangle _{ss} =\frac{2\left[ \tilde{\Gamma}_{+}\left( \tilde{\Gamma}%
_{+}-\tilde{\Gamma}_{-}\right) -2\tilde{\Gamma}_{0}^{2}+4\mu \tilde{\Gamma}%
_{0}\epsilon _{y}\right] \mu \epsilon _{x}}{\left( \tilde{\Gamma}_{+}+\tilde{%
\Gamma}_{-}\right) \left[ \tilde{\Gamma}_{+}\left( \tilde{\Gamma}_{+}-\tilde{%
\Gamma}_{-}\right) -4\mu\tilde{\Gamma}_{0}\epsilon _{y}+8\mu^2\epsilon _{y}^{2}%
\right] +8\mu^2\left( \tilde{\Gamma}_{+}-\tilde{\Gamma}_{-}\right) \epsilon
_{x}^{2}}.
\label{imas+}
\end{equation}

Inserting Eqs. (\ref{so}-%
\ref{imas+}) into Eq. (\ref{campototal}), and equating the real and imaginary parts, 
we get the following set of equations,
\begin{eqnarray}
E_{in} &=&\epsilon _{x} +\frac{\left( N-1\right) }{2V\mu }\left( \tilde{\Gamma}%
_{+}-\tilde{\Gamma}_{-}\right) {\rm Im}\left( \langle \sigma
_{+}\rangle _{ss}\right)  \label{stat11b} \\
0 &=&\epsilon _{y} -\frac{\left( N-1\right) }{2V\mu }\left( \tilde{\Gamma}%
_{+}+\tilde{\Gamma}_{-}\right) \left[ {\rm Re}\left( \langle
\sigma _{+}\rangle _{ss}\right) +\tilde{\Gamma}_{0}\langle \sigma
_{0}\rangle _{ss}\right] ,  \label{stat21}
\end{eqnarray}
for the components $\epsilon _{x}$ and $\epsilon _{y}$, and whose solutions determine $E_{ss}=\sqrt{
\epsilon _{x}^2+\epsilon _{y}^2}$ and $\Phi=\arctan\left(\epsilon _{y}/\epsilon _{x}\right)$.

By assuming, in particular, that the function $g^{2}(\omega )$ is symmetric around the atomic
transition frequency -- an even function with respect to $\omega _{0}$, $%
g^{2}(\omega _{0}-\omega )=g^{2}(\omega _{0}+\omega )$ --, 
the decay rates $\Gamma _{+}$ and $\Gamma _{-}$ become equal, so
the effective decay rates in Eq. (\ref{effectiverate}) reduce to $\tilde{%
\Gamma}_{\pm }=\frac{1}{2}\left( \Gamma _{0}\pm \Gamma _{+}\right) $ and $%
\tilde{\Gamma}_{0}=0$. In this case the output field does not depend on 
$\left\langle \sigma_0 \right\rangle$, from Eq. (\ref{stat21}) we have  
$\epsilon_{y} =0$, and the unique physical solution occurs for $\Phi = 0$ as in the case of a 
structureless reservoir. Nevertheless, the decay rates depend on the input field,  
so the relation between input and output field amplitudes simplifies to%
\begin{equation}
E_{in} = E_{ss} \left[ 1+\frac{N-1}{V}\frac{\Gamma
_{+}\left( E_{in}\right) }{\Gamma _{0} + \frac{16\mu ^{2}E_{ss}^{2}}{\Gamma _{0}+\Gamma _{+}\left( E_{in} 
\right) }}\right],  
\label{eq2}
\end{equation}%
where we wrote explicitly the dependence on $E_{in}$. In the
case of a structureless reservoir (white noise), $\Gamma _{0}=\Gamma _{+}$, Eq.
(\ref{eq2}) reduces to the well known equation \cite{boni1,xiao} 
\begin{equation}
E_{in}=E_{ss}\left[ 1+\frac{\left( N-1\right) }{V}\frac{1}{\left(
1+8\mu^2 E_{ss}^{2}/\Gamma ^{2}\right) }\right] , \label{Ein}
\end{equation}%
whose quite simple input-output amplitude nonlinearity is due to the 
term proportional to $N-1$. In order to compare with previous works \cite{boni1}, 
using the authors' notation, we write Eq. (\ref{eq2}) as
\begin{equation}
y=x +\frac{2C(y)x}{D(y)+x^2},  \label{bis2}
\end{equation}
where we have defined the input and output fields as $y=\sqrt{8}\mu
E_{in}/\Gamma_{0}$ and $x=\sqrt{8}\mu E_{T}/\Gamma_{0}$, respectively, $%
C(y)=((N-1)/V)(\Gamma_{0} +\Gamma_{+})\Gamma_{+}/4\Gamma_{0}^{2} $ and $D(y)=
(\Gamma_{0} +\Gamma_{+})/2\Gamma_0 $.
For a structureless vacuum we have $C(y)=C=(N-1)/V$, $D(y)=1$, so Eq. (\ref{bis2})
reduces to 
\begin{equation}
y=x + \frac{2Cx}{1+x^2}, \label{bis3}
\end{equation}
that is identical to the well known result obtained in \cite{boni1}. 
Equation (\ref{bis3}) displays a bistable behavior for $C>4$, and for $C\gg 1$, 
the range for input field, allowing three solutions, lies in the interval 
$\sqrt{8C} < y < C $. Now, regarding Eq. (\ref{bis2}), the dependence of 
$x$ on $y$, is obtained by inverting the expression and solving the 
cubic equation, which, for a structured environment, has its coefficients 
depending nonlinearly on the input field amplitude. So, the range 
of values for a bistable solution is determined 
by the density of particles and by the proper input field.

For a more general physical situation, when $g(\omega)$ is not symmetric around 
the atomic frequency $\omega_0$, all terms in Eq. (\ref{campototal})
contribute to the output field. In this case $\tilde{\Gamma}_0 \neq 0$ and an 
imaginary component $\epsilon_y$ arises in the output field. Thus, in contrast with the case of 
a symmetric $g(\omega)$ (or just being flat), the asymmetric structure of modes induces 
a relative phase in the output field, and both, amplitude and phase, display a bistable 
behavior. We illustrate below the input-output field dependence for two reservoirs models.
%
\subsection{Example 1: weighted Lorentzian shape}
%
We assume a structured reservoir having a weighted Lorentzian shape for the frequency 
distribution \cite{ref6,ref8},
\begin{equation}
g^{2}(\omega )=\frac{\Gamma _{0}}{2\pi }\left[ \beta +(1-\beta )\frac{\gamma
^{2}}{(\omega -\omega_r)^{2}+\gamma ^{2}}\right] ,  \label{density}
\end{equation}
with $\omega_r$ is a characteristic frequency of the reservoir and the 
parameter $0\leq $ $\beta \leq 1$. In Eq. (\ref%
{density}), the first term in the brackets
represents a background vacuum (white noise) whereas the second term
represents the structured vacuum (colored noise), assumed to have a
Lorentzian shape of width $\gamma $; the parameter $\beta $
interpolates between the two limiting cases respectively. The effective decay rates are
\begin{eqnarray*}
\tilde{\Gamma}_{-} &=&\Gamma _{0}\frac{\left( 1-\beta \right) }{4}%
\left[ -\frac{\gamma ^{2}}{\gamma ^{2}+\left( 2\mu E_{in}/\hbar+\eta \right) ^{2}}+\frac{2\gamma ^{2}}{%
\gamma ^{2}+\eta ^{2}}-\frac{\gamma ^{2}}{\gamma ^{2}+\left( 2\mu E_{in}/\hbar-\eta \right)
^{2}}\right]  \\
\tilde{\Gamma}_{0} &=&\Gamma _{0}\frac{\left( 1-\beta \right) }{4}%
\left[ -\frac{\gamma ^{2}}{\gamma ^{2}+\left( 2\mu E_{in}/\hbar+\eta \right) ^{2}}+\frac{\gamma ^{2}}{%
\gamma ^{2}+\left( 2\mu E_{in}/\hbar-\eta \right) ^{2}}\right]  \\
\tilde{\Gamma}_{+} &=&\Gamma _{0}\left\{ \beta +\frac{\left(
1-\beta \right) }{4}\left[ \frac{\gamma ^{2}}{\gamma ^{2}+\left( 2\mu E_{in}/\hbar+\eta
\right) ^{2}}+\frac{2\gamma ^{2}}{\gamma ^{2}+\eta ^{2}}+\frac{\gamma ^{2}}{\gamma ^{2}+\left(
2\mu E_{in}/\hbar-\eta \right) ^{2}}\right] \right\} 
\label{lorentzrates}
\end{eqnarray*}%
where $\eta =\omega _{r}-\omega _{0}$ is the detuning between atomic frequency
and the reservoir characteristic frequency. Worth to note that while $\tilde{\Gamma}_{-}$ 
and $\tilde{\Gamma}_{+}$ are even functions regarding the change $\eta \longrightarrow -\eta$,
$\tilde{\Gamma}_{0}$ is an odd function. The existence of an input-output phase difference 
depends on a nonzero $\tilde{\Gamma}_{0}$, and when one changes the sign of $\eta$, 
also that phase changes sign.  

\begin{figure}[tbp]
\includegraphics[angle=-90,scale=0.7]{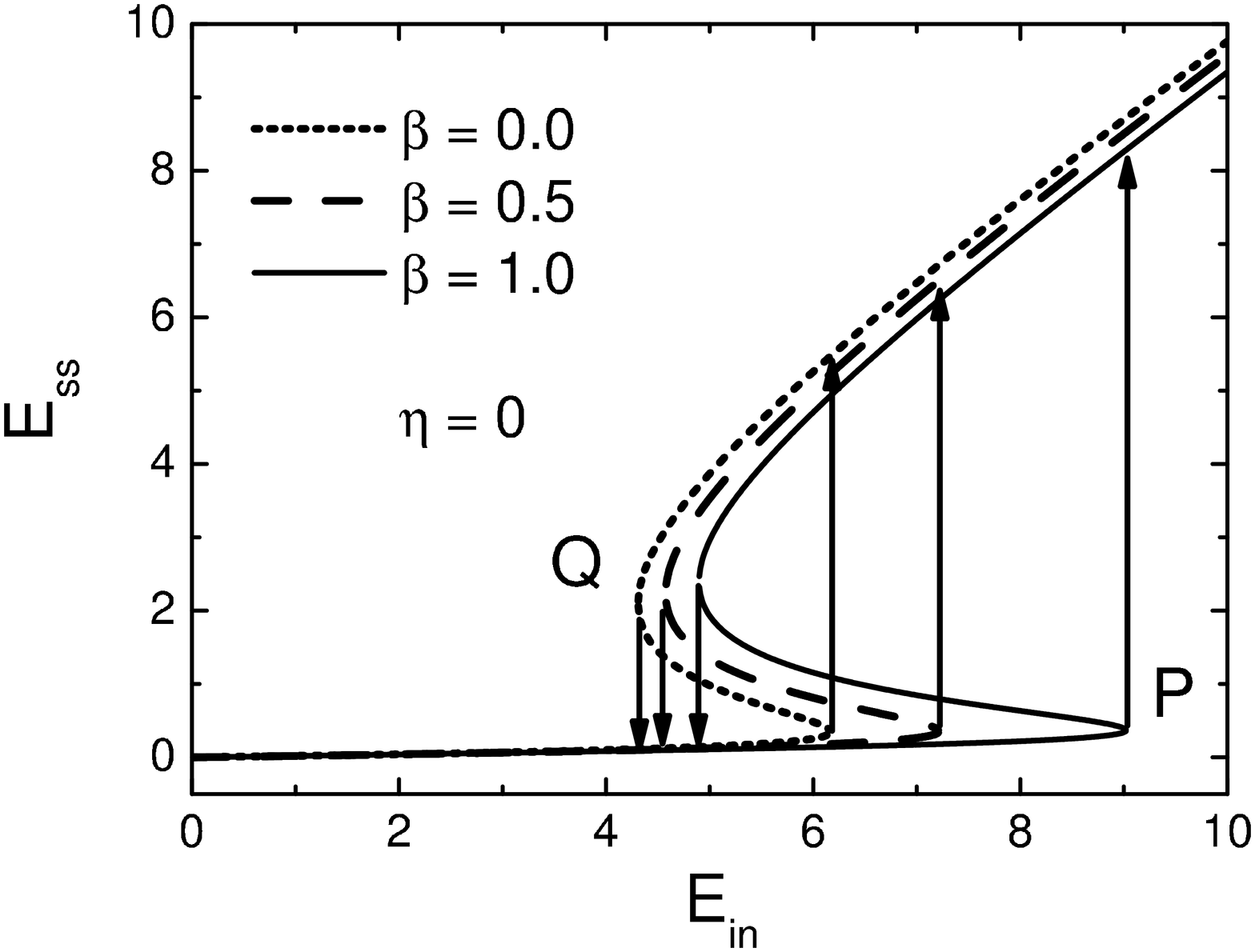}
\caption{Output versus input field amplitudes for $\eta = 0$, $N=50$, $\protect\gamma %
/\Gamma_{0} =20$ and different values of $\beta $. The solid line corresponds to an structureless 
reservoir, $\protect\beta =1$. The parameters are dimensionless.}
\label{f1}
\end{figure}

\begin{figure}[tbp]
\includegraphics[scale=0.7]{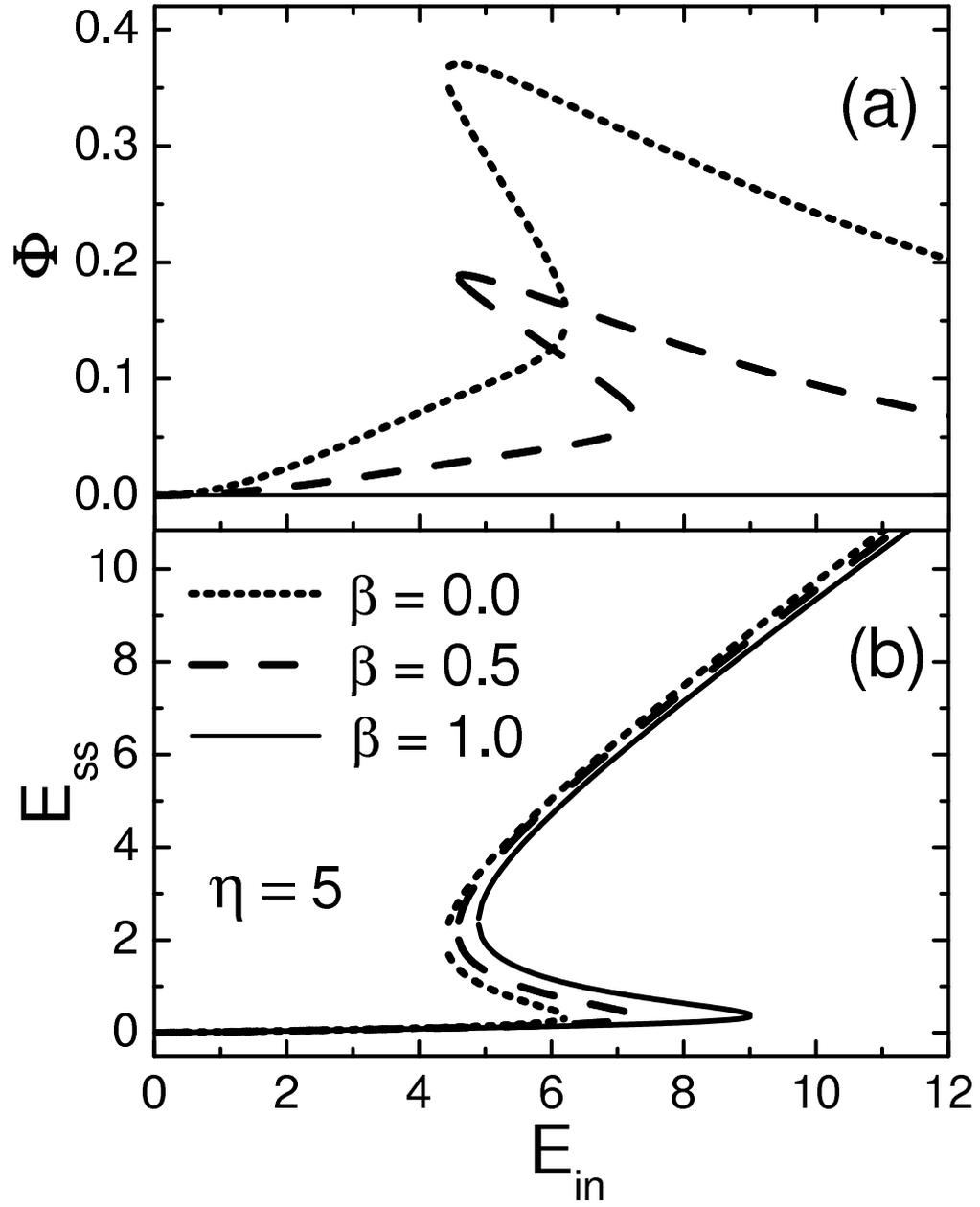}
\caption{Output (a) phase difference $\Phi$, in radians, and (b) amplitude $E_{ss}$, versus input amplitude for $\eta = 5$, $N=50$, $\protect\gamma %
/\Gamma_{0} =20$ and different values of $\beta $. The solid line corresponds to a structureless
reservoir, $\protect\beta =1$. The parameters are dimensionless. }
\label{f2}
\end{figure}

\begin{figure}[tbp]
\includegraphics[scale=0.7]{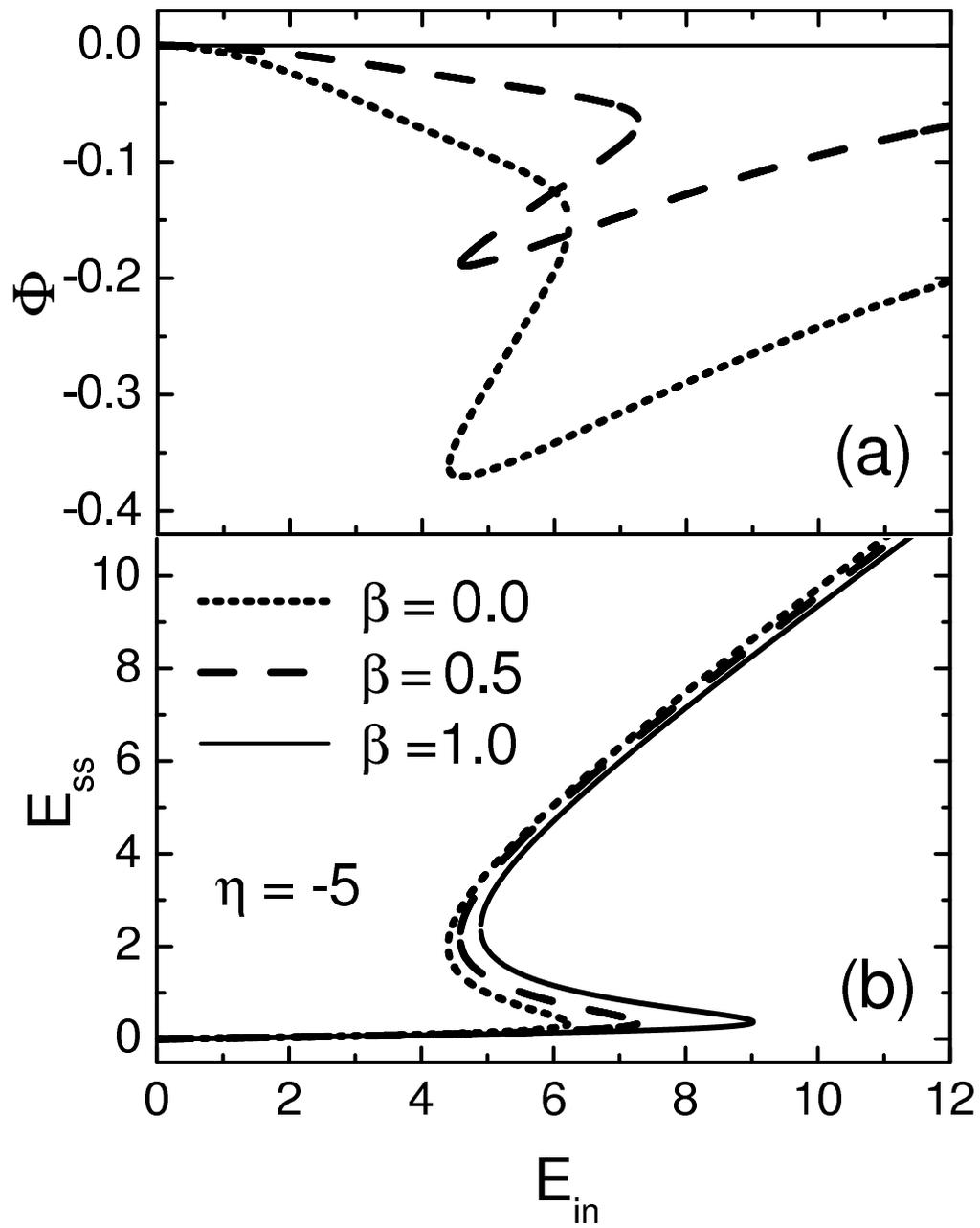}
\caption{The same as in Fig. \ref{f2} but for $\eta= - 5$.}
\label{f3}
\end{figure}

In Figs. \ref{f1} we plotted the output versus input amplitudes for three values of $\beta$, 
a detuning $\eta=0$, and $N=50$. In this case $\epsilon_y =0$ and $\epsilon _x =E_{ss}$, 
so the phase difference is $\Phi=0$. In both cases, $\beta =0.5$
and $\beta =0.0$ we observe a variation in the distances between the switching points, $P$ and $Q$, that
indicate the location of the lower and upper branches of the S-shaped curve. Comparing 
to the structureless reservoir, $\beta =1.0$ (solid line), there is a reduction in the range of values of
$E_{in}$ where the bistability occurs. We also observe that the deviations from the white noise 
curve (solid line) is more pronounced at the switching points ($P$) from lower to upper branches.
As the effective decay rate diminishes with the increase of the input field, less energy 
is transfered from the atomic sample to the reservoir, thus the energy goes through 
the sample carried by the output field.
We now consider the detuning $\eta=5.0$ and draw in Fig. \ref{f2}a and \ref{f2}b 
the output phase and amplitude, respectively, showing bistability in both. 
In Fig \ref{f2}a we observe that the phase difference is not 
null because of the emergence of the component $\epsilon_y$. The variation of $\Phi$ is more 
pronounced the more the mode distribution ($\beta =0.5$ and $\beta =0.0$) deviates 
from the white noise ($\beta = 1.0$). The phase $\Phi$ goes to zero for large values 
of the input amplitudes because the effective decay rate $\tilde{\Gamma}_{0}$ goes to zero.
Regarding the $\beta < 1$ cases, we note that amplitude and phase have the same switching points,
and the P - Q distance is reduced. By admitting a negative detuning $\eta=-5.0$ the input-output 
amplitude relation does not change, however, as can be seen in Fig.\ref{f3} the bistable behavior of the
phase changes by a sign inversion.
%
\subsection{Example 2: photonic band gap}
%
Here we adopt a simple reservoir structure used to analyze the resonance fluorescence 
phenomenon in a photonic band gap \cite{ref7}, where it is assumed that there is a
discontinuity at specific frequencies of the photonic density of modes $%
g^{2}(\omega )$, although it is constant over spectral regions in the
dressed atomic frequencies; so
\begin{equation}
g^{2}(\omega )=\left\{ 
\begin{array}{c}
\frac{\Gamma _{1}}{2\pi },\qquad \omega <\omega _{0} \\ 
\\ 
\frac{\Gamma _{2}}{2\pi },\qquad \omega \geq \omega _{0}.%
\end{array}%
\right.
\end{equation}%
In this case, at each dressed frequency, i.e. $E_{in}\neq 0$, the
decay rates are $\Gamma _{-}(\omega _{0}-2\mu E_{in}/\hbar )\equiv \Gamma
_{1}$ and $\Gamma _{0}\left( \omega _{0}\right) =\Gamma _{+}\left( \omega
_{0}+2\mu E_{in}/\hbar \right) \equiv \Gamma _{2}$, and the effective decay
rates become $\tilde{\Gamma}_{0}=\tilde{\Gamma}_{-}=\left( \Gamma
_{2}-\Gamma _{1}\right) /4$\ and $\tilde{\Gamma}_{+}=\left( \Gamma
_{1}+3\Gamma _{2}\right) /4$. 

\begin{figure}[tbp]
\includegraphics[scale=0.7]{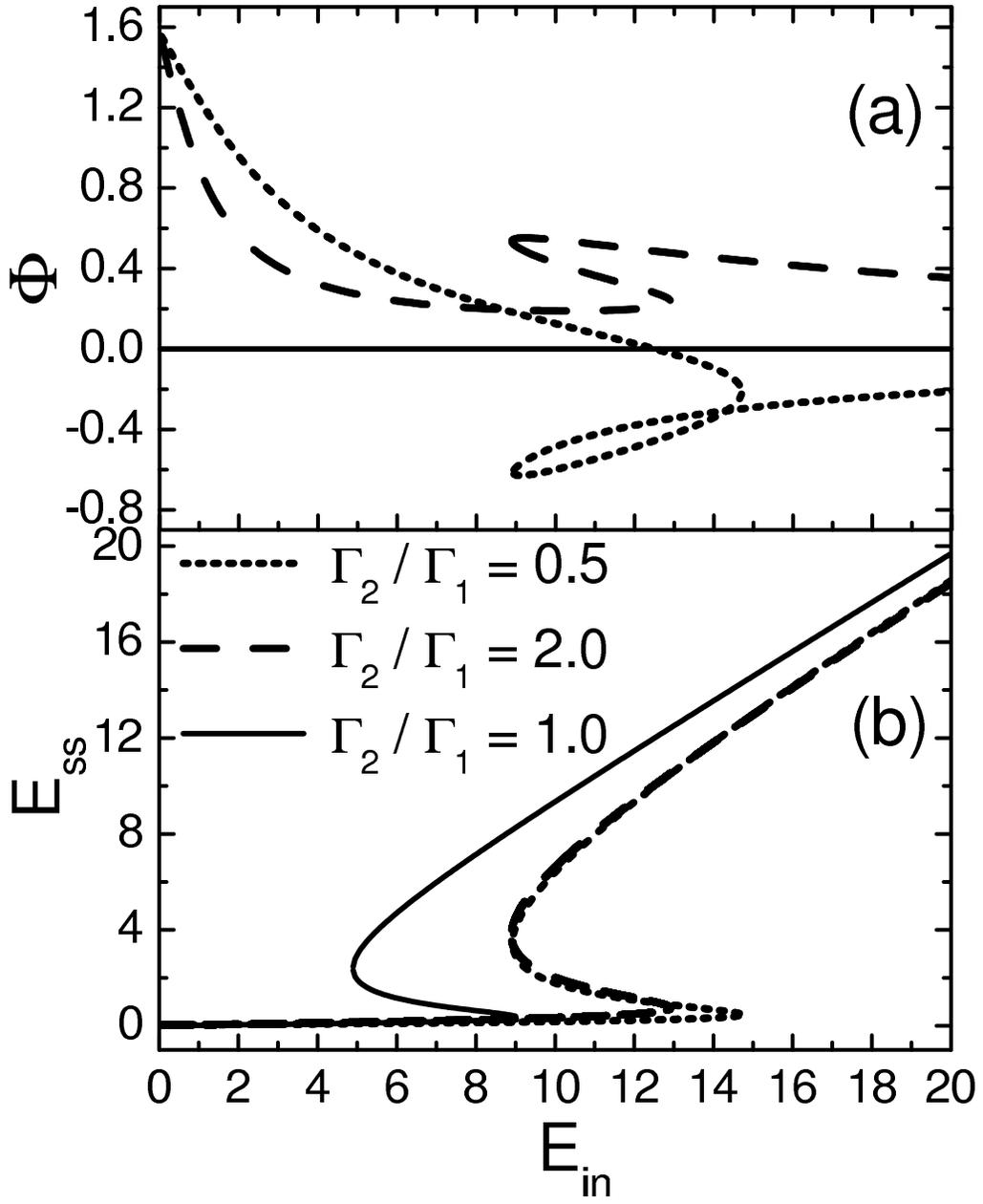}
\caption{Output (a) phase difference $\Phi$, in radians, and (b) amplitude $E_{ss}$, versus input field amplitudes for $N=50$, and different
values of $\Gamma _{2}/\Gamma _{1}$. The solid line corresponds to a structureless
reservoir, $\Gamma _{2}/\Gamma _{1}=1$. The parameters are dimensionless.}
\label{f4}
\end{figure}

In Figs. \ref{f4}a and \ref{f4}b we plotted the output field
phase and amplitude, respectively, as a function of the input amplitude, for some ratios 
of $\Gamma _{2}/\Gamma _{1}$. We set $N=50$, and all parameters are dimensionless. In Fig \ref{f4}a, 
as in the previous example, we observe a change in the amplitude input range for bistability. 
Also, the switching point from lower to upper branch is more sensitive to the ratio $\Gamma _{2}/\Gamma _{1}$. 
Regarding the phase, from Fig. \ref{f4}b we observe that, depending on the ratio $ \Gamma _{2}/\Gamma _{1}$, 
at the region of bistability the phase difference may change sign and also 
can display a more complex behavior (a loop) than an S-shaped form. For large values of the input 
amplitude, from Eqs. (\ref{so}-\ref{stat21}), we get $\epsilon_{x} \cong E_{in} $, $\epsilon_y$ attains, 
asymptotically, the constant value $(N-1)\Gamma _{2}(\Gamma _{2}-\Gamma _{1})/4V\mu(\Gamma _{1}+\Gamma _{2})$, 
and the phase $\Phi$ goes to zero with a sign that depends on the difference $\Gamma _{2}-\Gamma _{1}$. 
For larger values of  $E_{in}$ the phase difference $\Phi$ becomes proportional to the jump (discontinuity) 
in the structure of modes. 
%
\section{Summary and conclusions}
%
We presented a study of optical bistability using the mean-field approximation by considering a system of $N$
two-level atoms interacting with structured reservoirs. Our approach consisted in dressing the
atoms collective operators with the classical input field and coupling them to the structured
reservoir. The dynamical system is described by a master equation which contains 
extra terms (compared to that obtained from a 
structureless reservoir) resembling those present in the master equation,
derived under the influence of a squeezed reservoir. The master equation contains effective decay rates, 
associated to each dressed frequency, that depend on the frequency and intensity 
of the input field. Adopting the mean-field approximation, we deduced a single particle 
effective Hamiltonian containing extra nonlinear terms which are absent in the 
case of a structureless reservoir. In the stationary
regime of the atomic and laser field system, we analyzed the relation
between the input and output fields, observing that they are related in a nonlinear form
and present typical bistable behavior. For a structured reservoir the
S-shaped curve is present in the amplitude, but not necessarily in the phase. 
This theoretical result compares with a similar treatment given in \cite{tanas} where the authors analyzed the
resonance fluorescence and absorption spectra of a single atom and in which a similar phase
dependence in the system dynamics occurs. However, both the resonance
fluorescence and the absorption spectra have no phase dependence. Our results indicate 
that the presence of the induced phase shift in the output field of an N-atom system 
could be an interesting probe about the nature of the reservoir.  	
We have considered in details the case of resonance between atoms and the input field frequencies
and verified that the output phase shift appears 
for reservoirs having an asymmetric structure of modes. 
We presented two illustrative examples of
reservoirs: (1) a mixing of white noise and Lorentzian shaped
frequency distribution and (2) a simplified photonic band-gap distribution of the
reservoir modes. We noted that the output field bistable behavior is
as sensible in the phase difference, or acquired phase, as for the
amplitude. That characteristic can be explored by using a Mach-Zehnder 
interferometer, with a phase shifter in one of the arms, 
where the output field phase can
be determined by measuring the difference of the pulses intensities at the 
interferometer output ports. By slightly changing the input field $E_{in}$ at the
transition values in the instability region, it should be noted, as a
response, the occurrence of sudden and discontinuous changes in the difference between the
photocurrents at the two exit ports of the interferometer. On the other 
hand, by engineering reservoirs and atomic samples one could use the output amplitude 
and phase to design optical control devices. Besides, we will address, in a future work, 
the influence of structured reservoirs on bistability, in the dispersive regime ($\delta\neq 0$). 
\bigskip
%
\acknowledgments{GAP, ACY, and SSM, acknowledge financial support from FAPESP (S\~ao Paulo, SP,
Brazil), through the INCT-IQ, and  CNPq (DF, Brazil).} 
%



\newpage

\bigskip

\end{document}